\documentclass[12pt, paper, twoside]{JHEP3}
\usepackage{amsmath}
\usepackage{epsfig}

\renewcommand{\figurename}{Fig.}

\newcommand{\Lapl}{\vec{\partial}^{\, 2}}
\newcommand\bm\boldsymbol

\newcommand{\cA}{\mathcal {A}}
\newcommand{\tcA}{\tilde{\mathcal {A}}}
\newcommand{\cM}{\mathcal{M}}

\newcommand{\Kop}{\mathcal{K}}
\def\al{\alpha}

\def\de{\delta}

\def\lam{\lambda}

\def\half{\mbox{\small $\frac{1}{2}$}}

\title{High energy gravitational scattering:\\ a numerical study}
\author{Giuseppe Marchesini \\
  Universit\`a di Milano-Bicocca and INFN, Sezione di Milano-Bicocca\\
  Milano 20126, Italy\\ and\\
  Theory Division, CERN, Geneva, Switzerland \\
  E-mail: \email{Giuseppe.Marchesini@mib.infn.it} }

\author{Enrico Onofri\\ Universit\`a di Parma and INFN, 
 Gruppo Collegato di Parma\\
 Parma 43100, Italy\\
 E-mail: \email{Enrico.Onofri@unipr.it}}

\abstract{ The $S-$matrix in gravitational high energy scattering is
  computed from the region of large impact parameters $b$ down to the
  regime where classical gravitational collapse is expected to
  occur. By solving the equation of an effective action introduced by
  Amati, Ciafaloni and Veneziano we find that the perturbative
  expansion around the leading eikonal result diverges at a critical
  value signalling the onset of a new regime. We then discuss the main
  features of our explicitly unitary $S-$matrix down to the
  Schwarzschild's radius $R=2G\sqrt{s}$, where it diverges at a
  critical value $b\simeq 2.25\, R$ of the impact parameter.  The nature
  of the singularity is studied with particular attention to the
  scaling behaviour of various observables at the transition. The
  numerical approach is validated by reproducing the known exact
  solution in the axially symmetric case to high accuracy.}


\preprint{CERN-PH-TH/2008-042}

\begin{document}

\section{Introduction}\label{sec:introduction}

We discuss in this paper the high-energy quantum string-gravity
scattering by using the string formulation of Amati, Ciafaloni and
Veneziano \cite{ACV,ACV07}. Here we  concentrate on the study of
the high-energy elastic $S$-matrix for two colliding strings in the
limit in which the string length $\lam_s$ is negligible with respect
both to the gravitational Schwarzschild radius $R$ and to the impact
parameter $b$ of the the two colliding strings
\begin{equation}\label{eq:limit}
\frac{R}{b}\to0\,, \quad\frac{\lam_s}{b}\to0\,,\qquad 
R=2G\sqrt{s}\,,\quad \lam_s=\sqrt{\al'\hbar}\,,
\end{equation}
where $\sqrt{s}$ is the center of mass energy. In this
point-like-limit $d\!=\!4$ we rely on the observation that, because of
the soft multi-loop strings amplitudes \cite{ACV}, the $S$-matrix
exponentiates, in terms of an eikonal function of order $Gs/\hbar$,
times a function of $R^2/b^2$.  In this limit the two colliding
strings can be approximated by two massless particles
\begin{equation}\label{eq:S-Matrix}
S\>=\>e^{\frac{i}{\hbar}\,\cA_{\rm cl}}\,,\qquad
\cA_{\rm cl} \> = \> 2\pi G\, s \,\tcA_{\rm cl}\,.
\end{equation}
Here $\cA_{\rm cl}$ is the classical value of an effective action
given by a functional of gravitational fields constructed in terms of
effective graviton vertices derived by Lipatov \cite{Lipatov} in the
study of the multi-Regge limit in (QCD and) gravity. The fields
involved in the effective action $\cA$ are the longitudinal and the
transverse components of the metric tensor. The two colliding
particles are described in terms of the energy-momentum tensor along
the two light cone directions $x^\pm=x^0\pm x^3$.

The classical region to study is the case when $R$ is small and it
approaches a critical value $R_{\rm crit}$ at a fixed values of
$b$. Here the critical value of $R$ is obtained by two approximations:
we neglect rescattering terms and we neglect the infrared singular
term of graviton interaction.  Within these two approximations one is
left with the critical behaviour
\begin{equation}\label{eq:arctich}
R^2\,\frac{\partial}{\partial R^2}\,\tcA_{\rm cl}(b,s) \sim (1- R/R_{\rm
crit})^{1/2}\>\Rightarrow\> \tcA_{\rm cl}\sim (1-R/R_{\rm crit})^{3/2} +
{\rm constant}\,.
\end{equation}
This critical behaviour has been derived in a modified model
exhibiting axial symmetry (where the singular $\vec{b}-$dipole
disappears) but here it will be reproduced in the actual case of a
dipole at a distance $\vec{b}$.

The paper is organized as follows. First we introduce the model
(sec.~2) and explain the two relevant approximations for the dipole
picture with impact parameter $\vec{b}$. Then we explain the solution
for $R\le R_{\rm crit}$ and discuss the search of the critical index
(sec.~3). We then discuss the results and the behaviour of several
observables near critical region with $R\le R_{\rm crit}$.  Here we
discuss also the region $R>R_{\rm crit}$ in the axial case, which is
not solvable by our iteration technique.  We end with some
conclusions.

\section{The model}\label{sec:model}
Here we recall the essential points of the effective action introduced
in \cite{ACV,ACV07} which are needed to describe our numerical study
and we describe the two basic simplifications that are performed to
give
\begin{equation}
\label{eq:Act}
\begin{split}
\frac{\cA}{2\pi G s}
&=\int d^2x\>\Big( 
a_+\,t^+ +a_-\,t^- \>-\>\half\vec{\partial} a_+ \cdot \vec{\partial}a_-
\>+\>\half(\pi R)^2\left(-(\Lapl \phi)^2
+2\phi\, S_{a_+a_-}\right)
    \Big)\\[3mm]
&    S_{f,g}(\vec{x}) \equiv
    \Lapl f(\vec{x})\;
    \Lapl g(\vec{x})\> - 
    \partial_i\partial_j f(\vec{x})\;
    \partial_i\partial_j g(\vec{x})\,.
\end{split}
\end{equation}
Here $a_\pm(\vec{x})$ are the longitudinal graviton components (dotted
lines in Fig.~1). Upon factorization the light-cone delta-function
$\de(x^\mp)$ respectively, they depend only on the transverse
variables $\vec{x}=(x_1,x_2)$. In the same way $t_\pm(\vec{x})$
correspond to the longitudinal components of the energy-momentum
tensor associated to the two colliding particles at impact parameter
$\vec{b}$
\begin{equation}
\label{eq:tpm}
\begin{split}
t_\pm(\vec{x})=\de^2(\vec{x}\mp\half\vec{b})\,.
\end{split}
\end{equation}
The field $\phi(\vec{x})$ corresponds (upon factorizing the light-cone
function $\theta(x^+x_-)$) to the transverse graviton component (the
wavy line in Fig.~1a). 

\EPSFIGURE[t] {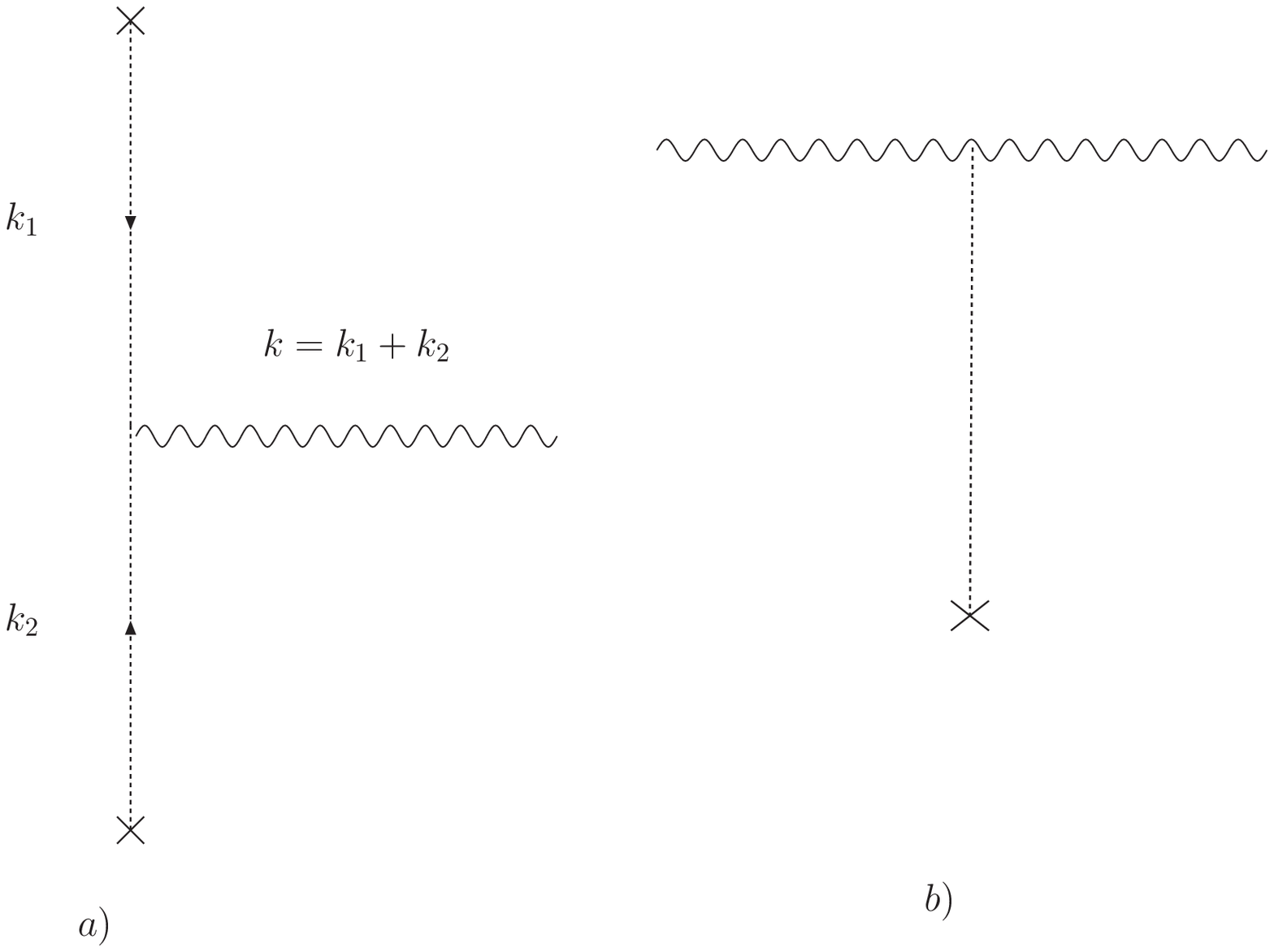, width = 12cm}{ (a) the scattering vertex
  between the two partons at distance $\vec b\,$.  The crosses are the
  external particles. Dotted lines are $a_\pm$ and wiggly lines denote
  $\phi$.~ (b) The rescattering of partons.
 \label{fig:Lipatov}}

The first two terms of $\tcA$ correspond to the eikonal exchange of
longitudinal gravitons (Coulomb scattering).  This part of the action
($R=0$) gives rise to the classical field 
\begin{equation}
\label{eq:sol0}
\begin{split}
&a^{(0)\rm cl}_\pm(\vec{x})=-\frac{1}{\pi}
\ln (\lam|\vec{x}\pm\half\vec{b}|)\,,
\end{split}
\end{equation}
with $\lam$  an infrared cutoff. The corresponding classical action 
is given by 
\begin{equation}
\label{eq:A0}
\begin{split}
\frac{\cA^{(0){\rm cl}}(b)}{2\pi Gs}=
-\frac{1}{\pi}\ln(\lam b)\,,
\end{split}
\end{equation}
which shows that the infrared cutoff $\lam$ is a divergent Coulomb
phase contribution which does not affect observables. It reproduces
Einstein's deflection angle $\sin\half\theta_{\rm cl} =
R/b$.

The other pieces of $\cA$ correspond to quantum correction
proportional to $R^2/b^2$. In particular the last term corresponds to
the effective vertex \cite{Lipatov} in Fig.~1a.
The form of $\cA$ in Eq.~\eqref{eq:Act} is based on two
simplifications.

\begin{itemize}
\item No rescattering: the effective vertex in Fig.~1b, the
  rescattering term, is neglected. In this case the dependence of the
  fields on the light-cone directions $x^+$ and $x^-$ of the two
  colliding particles can be factored into the two delta functions
  $\de(x^\pm)$. Dropping this ``trivial'' dependence, one is left with
  the fields in the effective action which depend only on the two
  transverse space components $\vec{x}$.

\item No infrared-singular transverse graviton component.  The
Lipatov vertex of Fig.~1a is given by
($\theta_{12}=\theta_{\vec{k_1}\vec{k_2}}$)
\begin{equation}\label{eq:V}
V_{\rm LLT}^{\mu\nu}(\vec{k_1},\vec{k_2},\vec{k})
\sim \frac{1}{\vec{k}^2} 
\Big(\sin^2\theta_{12}\,\epsilon_{\rm TT}^{\mu\nu}(\vec{k})-
\sin\theta_{12}\,\cos\theta_{12}\,\epsilon_{\rm LT}^{\mu\nu}(\vec{k}) \Big)\,,
\end{equation}
with $\epsilon_{\rm TT}$ and $\epsilon_{\rm LT}$ the two independent
transverse-traceless tensors of the graviton (see for example Eq.~2.5
of Ref.~\cite{ACV07}). For $\vec{k}\to0$ the first term
$\sin^2\theta_{12}/\vec{k}^2$ remains finite while the second one,
$\sin\theta_{12}\,\cos\theta_{12}/\vec{k}^2$ diverges.  In our study
we neglect the $\epsilon_{LT}$ transverse graviton component which
brings in infrared divergences.  The possible r\^ole of these singular
components and the cancellation of the infrared singularities in the
$S$-matrix has been discussed in \cite{ACV07} by using coherent states
similar to what is done in QED and QCD \cite{CCM}.

In the space representation the first term of the vertex \eqref{eq:V}
gives rise to the $2\phi\,S_{a_+,a_-}$ term in the action. Indeed
\begin{equation}
\label{eq:VTT}
F(\vec{k_1})\,G(\vec{k_2})\,
\sin^2\theta_{12}
\quad \Rightarrow\quad
S_{f,g}(\vec{x})\,,
\end{equation}
with $f(\vec{x})$ and $g(\vec{x})$ the Fourier transforms of the two
functions $F(\vec{k_1})$ and $G(\vec{k_2})$. The second term, infrared
singular, gives ($\epsilon_{12}\!=\!\epsilon_{21}\!=\!-1$)
\begin{equation}
\label{eq:VLT}
F(\vec{k_1})\,G(\vec{k_2})\,\sin\theta_{12}\cos\theta_{12}
\quad \Rightarrow\quad 
\vec{\partial}\,\partial_i f(\vec{x})\,\cdot\,
\vec{\partial}\,\partial_j g(\vec{x}) \,\epsilon_{ij}
\end{equation}
which for the free solution \eqref{eq:sol0}, with $f = a^{(0)\rm
  cl}_+,\; g = a^{(0)\rm cl}_-$, is proportional to $\vec{x}\wedge \vec
b$. The contribution from this infrared singular vertex is neglected
in Eq.~\eqref{eq:Act}.
\end{itemize}
\noindent
The equations of motion \eqref{eq:Act} are readily obtained by using
the fact that $\phi\,S_{a_+a_-}=a_\pm\,S_{\phi\, a_\mp}$ up to total
derivatives.  One has
\begin{equation}
\label{eq:eom}
\begin{split}
&\Lapl\,a_\pm(\vec{x})=-2\,\de^2(\vec{x}\pm\half\vec{b})
+2(\pi R)^2\,S_{\phi\,a_\pm}(\vec{x})\\
&(\Lapl)^2\,\phi(\vec{x})=-S_{a_+\,a_-}(\vec{x})
\end{split}
\end{equation}
If $a_\pm(x), \phi(x)$ is a solution, then
\begin{equation}
\label{eq:sym}
\tilde a_\pm(\vec{x})=a_\mp(\vec{b}-\vec{x})\,,\qquad
\tilde \phi(\vec{x})=\phi(\vec{b}-\vec{x})\,.
\end{equation}
is also a solution.  If the solution is unique (e.g. for small $R$)
then it is symmetric.  Perturbative corrections in ${R^2/b^2}$ to
Eq.~\eqref{eq:sol0} are finite. To first order the action is given by
\begin{equation}
\label{eq:A1}
\begin{split}
\frac{\cA^{\rm cl}(b)}{2\pi Gs}=
\frac{1}{\pi} \left(-\ln(\lam b)+\frac{3R^2}{8b^2}+\cdots\right).
\end{split}
\end{equation}
Here one has used \eqref{eq:Act} thus neglecting the transverse 
$\epsilon_{LT}$
graviton component. This latter has been computed in \cite{ACV07} and
gives an additional contribution $R^2/(8\pi b^2)$.

\section{Solution by an iterative algorithm}\label{sec:solut-an-iter}
In the small $R$ regime we solve Eq.\eqref{eq:eom} by iteration
($\Delta=\Lapl$)
\begin{eqnarray}
 \label{eq:relax}
  a_\pm(\vec{x}) &\leftarrow&(1-\omega) a_\pm(\vec{x}) - \omega\,
  \Delta^{-1}\,\left( 2\,\delta^2(\vec{x} \mp \vec{b}/2) - 2(\pi
  \,R)^2\, 
  S_{\phi,a_\pm}\right)\\ \phi(\vec{x})
  &\leftarrow&(1-\omega)\phi(\vec{x}) - \omega\, \Delta^{-2}\, S_{a_+,
  a_-}
\end{eqnarray}
($S_{f,g}$ as given in Eq.~(\ref{eq:Act})).  Here $\omega$ is a
``relaxation parameter'' which may help convergence; in a linear
context, the iteration converges if the r.h.s. has norm less
than one, using the relaxation parameter can be useful if all
eigenvalues have real part less than one.  In our non-linear context
the role of relaxation is not immediately clear. It is only used to
check that divergence of the iteration is not changed by a trivial
modification of the iteration mechanism. All results presented later
on correspond to $\omega=1$.

In the actual implementation of the algorithm, we substitute the Dirac
deltas with extended normalized Gaussian source terms 
\begin{equation}
  \label{eq:rho}
  \delta^2(\vec x \pm \vec b/2)\rightarrow \mathcal N \exp\left\{  
\frac{-(\vec x \pm \vec b/2)^2}{2\sigma}\right\}
\end{equation}
since dealing with smooth functions improves the numerical stability;
the Gaussian width $\sigma$ enters as a smearing parameter in the
calculation.
\subsection{Algorithmic details}\label{sec:algorithmic-details}
The calculation is organized as follows. We choose a finite
$2-$dimensional lattice of size $2L$ and lattice spacing $2L/N$,
typically $L=64, N=256$; we introduce the dual lattice $(k_1,k_2)$ and
define the basic differential operators in terms of $k_i$,
i.e. $\Delta \rightarrow -k_1^2-k_2^2$, etc. The r.h.s. of the
iteration map is then computed by going to Fourier space where
necessary and taking local products of fields in coordinate
space. This is done very efficiently for any $N$ by using FFTW
\cite{FFTW}, the best implementation of the fast Fourier transform to
our knowledge.  The main problem is to monitor the iteration and stop
it when it can be decided that we have convergence or divergence.  To
monitor the iteration we choose a specific component of the action,
namely the part proportional to $R^2$ ($\cA=2\pi Gs\,\tcA$)
\begin{equation}
  \label{eq:monitor}
  \cM = \half(\pi R)^2 \int d^2x \> \left(-(\Lapl 
\phi)^2+2\phi\, S_{a_+a_-}\right) \equiv \tcA_{\phi^2}+\tcA_{\phi\, a_+
a_-} = R^2\frac{\partial}{\partial R^2}\,\tcA\;.
\end{equation}
A second monitoring device is provided by the fact, already used in
\cite{ACV07}, that on the solutions one has
\begin{equation}
  \label{eq:Euler}
  2 \tcA_{\phi^2} + \tcA_{\phi\, a_+ a_-} = 0\,.
\end{equation}
\noindent
When $\mathcal M$ shows a variation less that a certain prefixed value
(say $10^{-6}$) then we stop the iteration and proceed to a more
stringent convergence check based on the analysis of the linearized
equation near the approximate solution \footnote{For $R>R_{\rm crit}$
  the iteration will diverge and the program will go in overflow
  unless we apply some strategy to early identify a divergent
  behaviour. A possible method is to monitor the {\sl curvature\/} of
  the function $\cM$ as a function of the iteration count.}.  This is
done as follows: let
\begin{equation}
  \begin{pmatrix}
    a_\pm\cr \phi
  \end{pmatrix} \leftarrow \mathcal K(a_\pm, \phi)
\end{equation}
denote the iteration map; near convergence let $\phi = \tilde\phi +
\delta\phi$, $a_\pm = \tilde a_\pm + \delta a_\pm$ where $\tilde\phi,
\tilde a_\pm$ represent the approximate solution obtained at the last
iteration. Then we have
\begin{equation}
  \label{eq:lineAr}
  \begin{pmatrix}
    \delta a_\pm\cr
\delta \phi
  \end{pmatrix} \leftarrow 
  \begin{pmatrix}
    \mathcal D_{11}&\mathcal D_{12}\cr
    \mathcal D_{21}&\mathcal D_{22}
  \end{pmatrix}
  \begin{pmatrix}
    \delta a_\pm\cr
    \delta \phi    
  \end{pmatrix}
\end{equation}
where the matrix $\mathcal D_{ij}(\tilde a_\pm, \tilde\phi)$ is
obtained by linearizing the equation and discarding quadratic terms in
$\delta a_\pm,\, \delta \phi$. The spectrum of $\mathcal D_{ij}$ is
then computed. If the spectrum is contained in the unit circle then we
conclude for convergence, otherwise for divergence\footnote{As an
option we may consider the eigenvalue with largest real part: if this
is larger than one we have divergence, otherwise convergence could be
achieved by a suitable choice of relaxation parameter.}. The linear
operator $\mathcal D$ contains partial differential operators and the
fields $a\pm,\, \phi$. Its largest eigenvalue, in absolute value, is
computed\footnote{Recent versions of Matlab$^\copyright$ provide a
user-friendly interface to the package.} by the Arnoldi algorithm
contained in the mathematical library ARPACK \cite{ARP}.

\subsection{The search for $R_{\rm crit}$}\label{sec:search-r_rm-crit}

Running the algorithm, we observe that the iteration converges at
small values of $R$, where we effectively obtain the perturbative
solution. However, increasing the parameter one finds that the
algorithm ceases to converge at a value of $R=R_{\rm crit}$
proportional to the impact parameter.

To  identify the value of $R_{\rm crit}$ where the transition
occurs we adopt a ``bisection method''. Namely we start from an
interval $[R_{min}, R_{max}]$ and set $R=\half(R_{min}+ R_{max})$; if
the iteration starting from $R$ is divergent we set
$R_{max}=R$ otherwise $R_{min}=R$; the process is replicated until
$R_{max}-R_{min}$ is less than a desired accuracy (typically
$10^{-5}$).

The value thus obtained is however affected by the presence of
``technical'' parameters whose impact on the calculation must be
carefully analyzed. The parameters are $L, N, \omega, \sigma$,
i.e. infrared and ultraviolet cutoffs, the relaxation parameter and
the width of the sources. On dimensional grounds one has
\begin{equation}
  \label{eq:FSS}
R_{\rm crit} = b\, F_N(\sigma/L^2, b/L, \omega)\,,
\end{equation}
and, in principle, the limit $\sigma\to 0, L\to \infty, N\to \infty$
should be taken. It turns out that the iteration scheme is not
sensitive to the relaxation parameter, hence we drop it from now on.
On the other hand the dependence on infrared ($L$) and ultraviolet
($\sigma$) parameters is quite accentuated and some fitting procedure
must be adopted in order to get the continuum infinite volume limit.
Dependence on $N$ on the other hand is very weak above $N\sim 100$, as
we argue in next paragraph.

\paragraph{Continuum limit and Finite Size Scaling.}
To estimate the $N$ dependence we performed a series of iterations
starting with different parameters but keeping the adimensional ratios
inside $F_N$ fixed. One finds that the results reach a flat plateau
very soon (above N=100 the numbers agree to 4 figures).  This is
surely due to the way differential operators are dealt with: any
finite--difference scheme would introduce a systematic error of
$O(N^{-n})$ - e.g. the roughest scheme corresponds to $\Lapl \to
2\sum_i(1-\cos(a k_i))/a^2$; using $k^2$ itself in Fourier space makes
the error {\sl exponentially small} in $N$.

Having settled the $N$ dependence, systematic errors coming from
$\sigma$ and $L$ should be addressed.  A long run on a lattice of
values for $b, \,\sigma$ and $L$ manifests a good scaling behaviour
(i.e. the plot of $R^2/b^2$ is acceptably smooth, hence showing
compatibility with a scaling behaviour).

\section{Results}\label{sec:results}
The main result regards the dependence of the critical value of $R$
with respect to the impact parameter. We ran the program for a
three-dimensional grid of values $\sigma \ll b \ll L)$ and we looked
for a best fit to the $F_N$ in Eq.~(\ref{eq:FSS}) assuming a simple
form
\begin{equation}
  \label{eq:1}
  F_N = \sqrt{1 + \alpha_1 \sigma / b^2 + \alpha_2 (b/L)^2 +
    \alpha_3 (b/L)^4 + \ldots\;}.
\end{equation}
This gives (see Fig.\ref{fig:dipole})
\begin{equation}
  \label{eq:slope}
  R_{\rm crit} \approx 0.445(1)\, b\;\;.
\end{equation}

\EPSFIGURE[t]{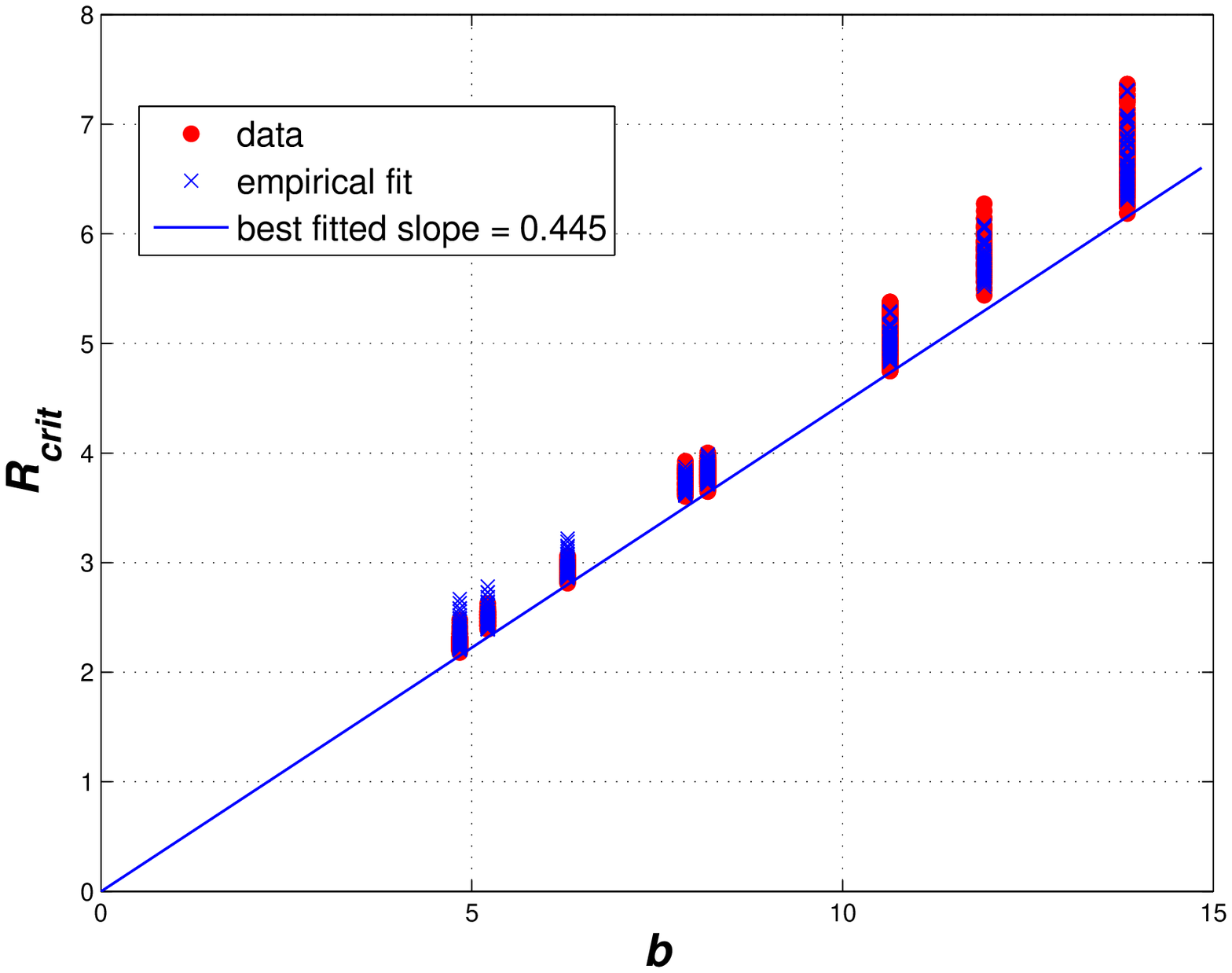, width=10.cm}{$R_{\rm crit}$
  vs. $b$ in the dipole model.\label{fig:dipole}} 

\noindent
The evidence from the fit is that the correct slope corresponds to the
envelope of the data from below, which can be interpreted in the
sense that ultraviolet and infrared cutoffs both make the system more
stable against collapse.

\paragraph{The axially symmetric case.}
A comparison with the result \cite{ACV07,VW} obtained in the case of
axial symmetry, namely $R_{\rm crit}/b = 2^{1/2}/3^{3/4}\approx 0.6204$
will support our analysis. In this case one of the sources is
uniformly distributed around an annulus of radius $b$, while the other
is localized at the center. The difference between this result and the
value in Eq.~\eqref{eq:slope} suggests that the critical value, due to
the non--linearity of the equations, may depend on the sources.  In
order to validate the accuracy of our numerical code, we applied it to
the study of this axially symmetric case. A fit conducted with the
same systematics as before gave the result 
\begin{equation}
  \label{eq:axial}
  R_{\rm crit} \simeq  0.6216
\end{equation}
which presents the same pattern as in Fig.~\ref{fig:dipole} and
reproduces the exact slope to an accuracy of $0.5\%$; this check makes
us confident on the accuracy of our code and allows us to estimate the
error in Eq.~\eqref{eq:slope} to less than 1\%.

\subsection{Critical behaviour}\label{sec:critical-behaviour}
In order to better understand the nature of the transition at
$R=R_{\rm crit}$ we shall now present some results about the critical
behaviour of certain observables. The main fact we derive from our
numerical data is that all observables that we examined have a scaling
behaviour near the transition which can be reproduced very accurately
by a \emph{square root singularity}. This fact supports the conclusion
that we are in presence of a genuine transition and not simply a
breakdown of the iteration scheme. The argument is as follows: the
iteration scheme represents an efficient way to sum up the
perturbative expansion in the parameter $\Kop=2(\pi R)^2$; as such the
iteration's convergence radius is regulated by the nearest singularity
in the complex $\Kop$ plane. Our analysis shows that the divergence of
the iteration scheme is caused by a singularity on the real line,
which must then correspond to a physical singularity.

\paragraph{Spectral properties.} 
Let's start with the spectrum of the linearized equation which is
used in monitoring the convergence of the iteration algorithm.  From
the data a scaling property emerges which appears to be rather
robust against variations of other parameters, namely the dependence
of the spectral radius against $R$. Let $\lambda_0$ denote the
spectral radius of the linearized equation: the plot of $1-\lambda_0$
as a function of $\sqrt{1-R/R_{\rm crit}}$ is reported in the next
picture (Fig.~\ref{fig:srs}) and it suggests a relation of the type
  \begin{equation}
    \label{eq:scaling}
    1 - \lambda_0  = C\,\sqrt{1 - R/R_{\rm crit}}
  \end{equation}
  with $C$ close to one.  

  \EPSFIGURE[t]{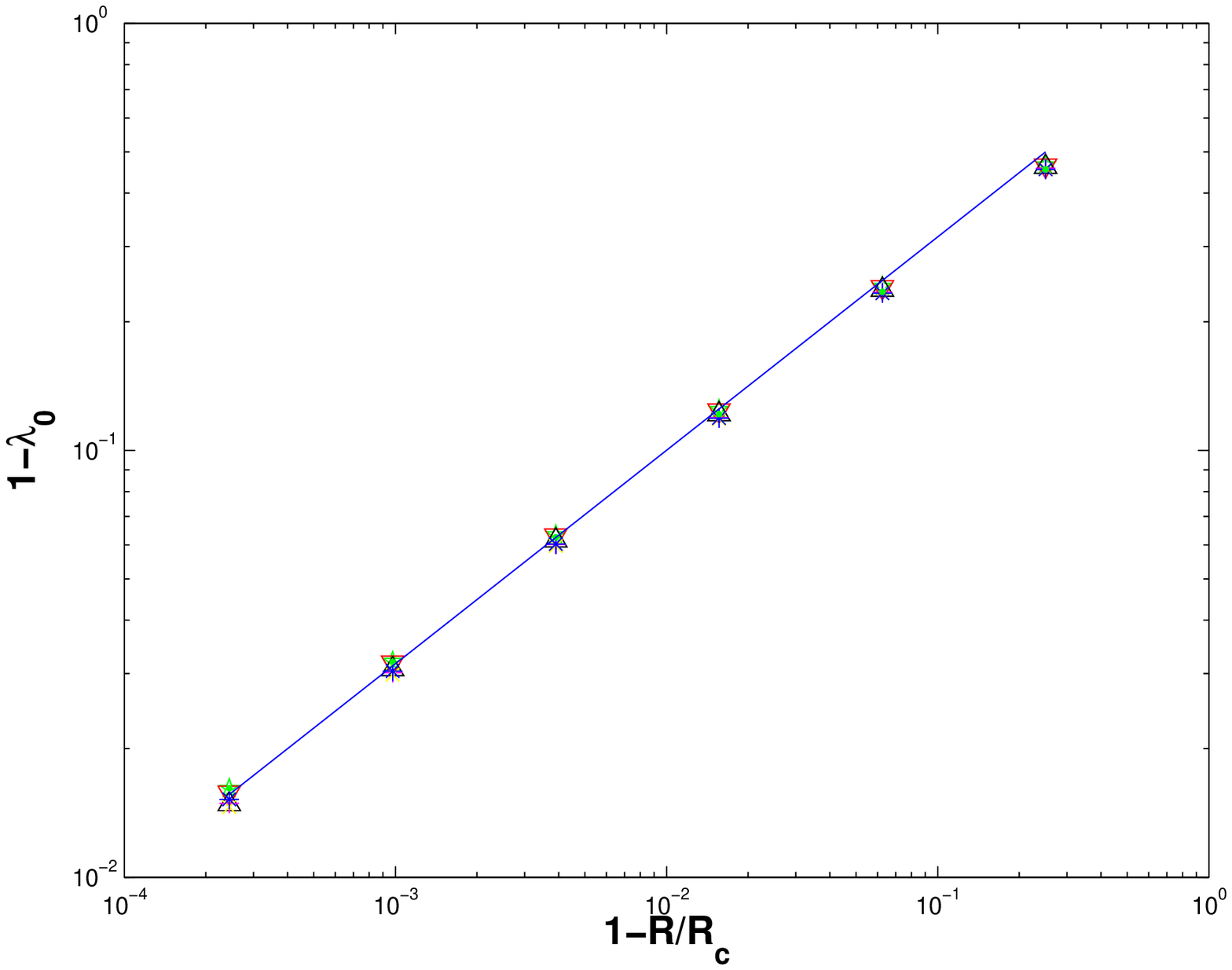, width=10.cm}{{\sl log-log} plot of the spectral
    radius as a function of $R$ in the dipole model.\label{fig:srs}}

\noindent
The plot reports {\bf data from different values of } $\boldsymbol b$
and it hints at the fact that there is universality in the relation
$\lambda_0 = \lambda_0(R/R_{\rm crit})$, i.e. the dependence on the
impact parameter is only through the function $R_{\rm crit}(b)$ and it
is quite insensitive to the various cutoff parameters.

A further investigation regards the region $R>R_{\rm crit}$ in the
case of axially symmetric sources. It is known \cite{ACV07,VW} that
above the transition the solution becomes complex. We computed the
spectrum of the linearized equation around the exact solution. The
spectral radius is compatible with the analytic continuation of
Eq.~(\ref{eq:scaling}), namely the largest eigenvalue gets an
imaginary part - see Fig.~\ref{fig:above}.

\DOUBLEFIGURE[t]{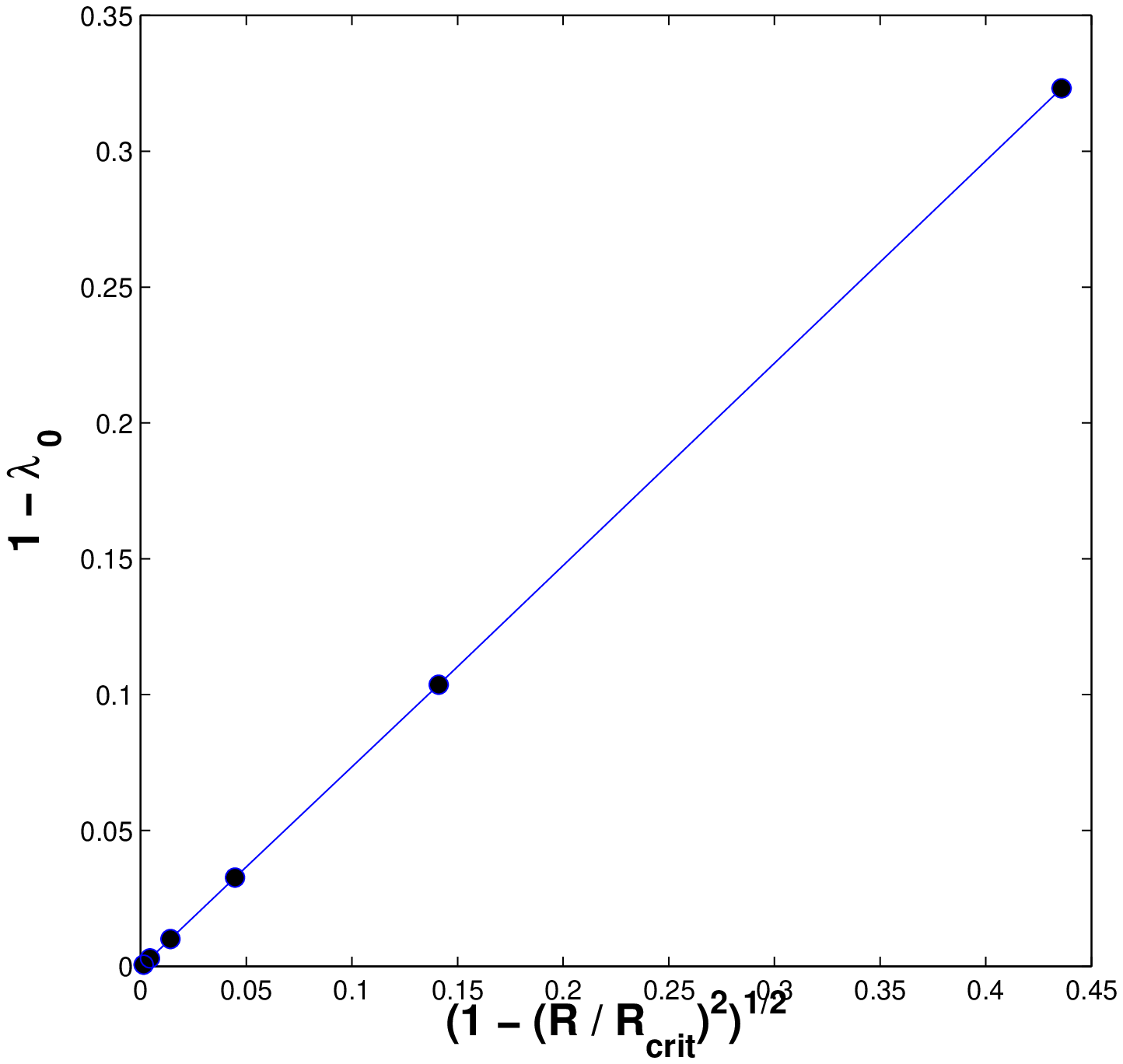, width=6.5cm}{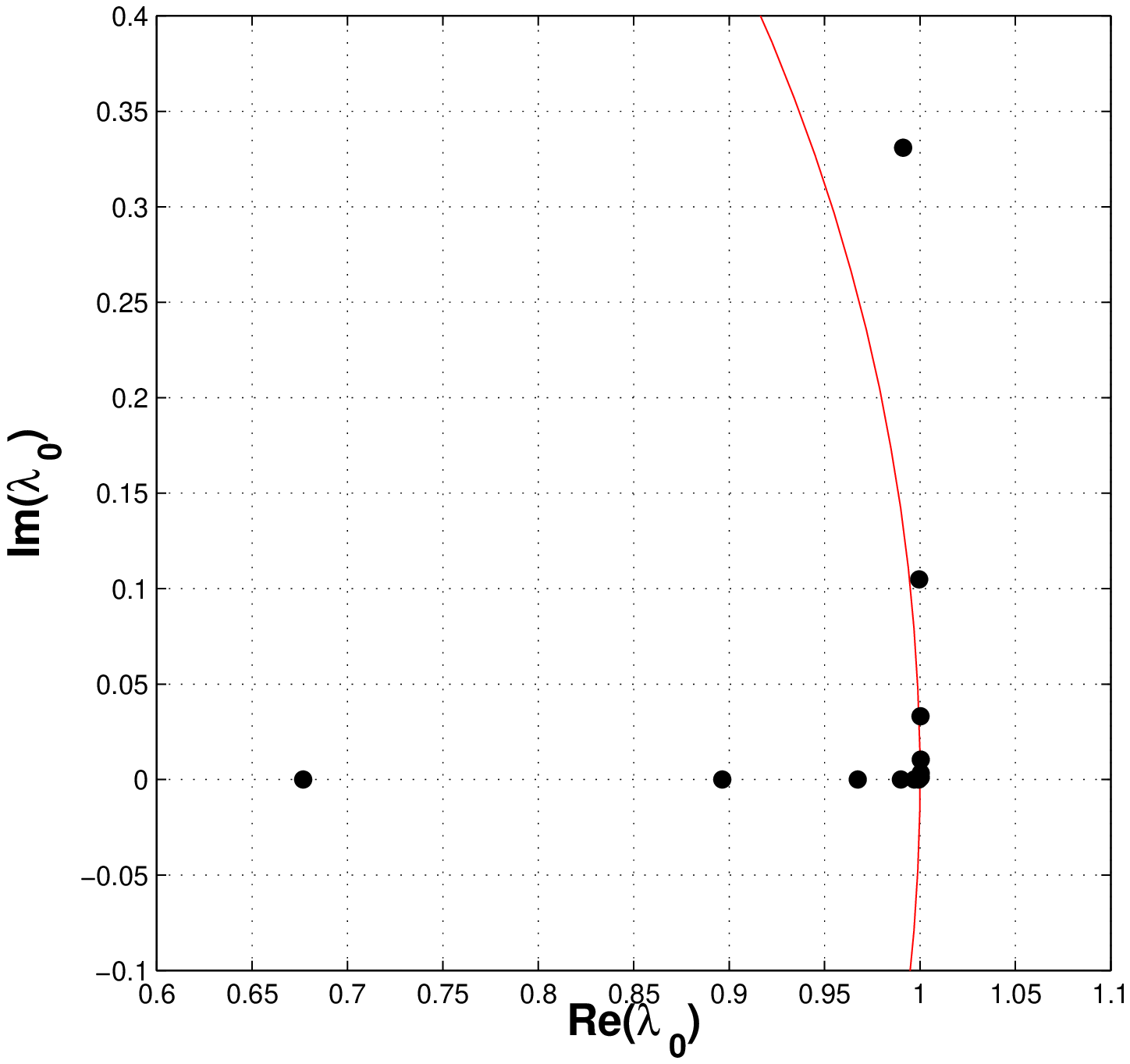,
  width=6.5cm} {The scaling of the leading eigenvalue in the axial
  symmetric case below criticality} {The leading eigenvalue above
  $R_{\rm crit}$ in the axial symmetric case above
  criticality \label{fig:above}}

\paragraph{Classical Action, $h_{T T}$ and  $h_{L T}$.} 
Near criticality we examined other observables, namely global
quantities like
\begin{itemize}
\item the action piece $\cM$ as in Eq.~\eqref{eq:monitor};
\item $\int\,d^2x \; h^2_{LT}$, which does not enter the action in the
  present approximation;
\item $h_{TT} = \Delta\phi$, integrated over a finite region $\propto
  b$.
\end{itemize}
As expected, we observed the same square root behaviour for $\cM$
\begin{equation}
  \label{eq:Mscaling}
  \cM = R\frac{\partial}{\partial R} \tcA(b,R) \sim \sqrt{1 - R/R_{\rm crit}}
\end{equation}
and for other cases - see Fig.~\ref{fig:action}. By integrating over
$R$, one gets Eq.~\eqref{eq:arctich}.

\EPSFIGURE[t]{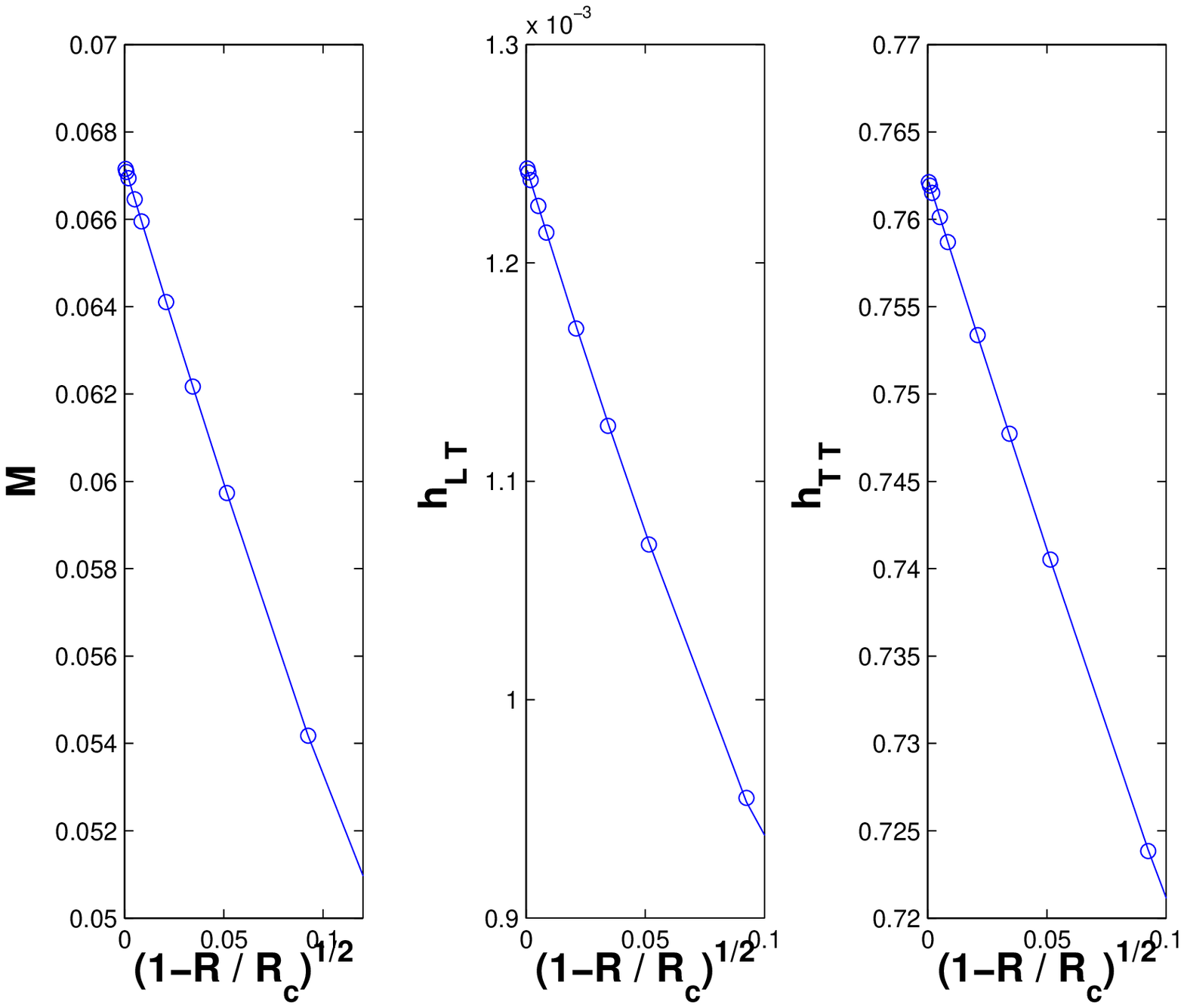, width=12.cm} {Scaling behaviour of $\cM$,
  $h_{T T}$ and $h_{L T}$ near Local observables. \label{fig:action}}

\section{Conclusions} 

We have neglected the string-size effects for $R\,,b\gg \lambda_s$
(with $\lambda_s$ the string effect) and studied the region without
rescattering and without trace infrared effects for $R$ small compared
to $b$ in \eqref{eq:Act}. We arrived at the critical region in
\eqref{eq:Mscaling} with the square root behaviour ensured for various
decades. This result nicely agrees with that obtained in the axially
symmetric case \cite{ACV07,VW}.
The fact that at the transition an eigenvalue 1 appears in the
spectrum of the linearized equation shows that the solution (if any)
at $R>R_{\rm crit}$ is unstable against small perturbations, hence
features that have been considered irrelevant in the derivation of
Eq.~\eqref{eq:Act} may come back as relevant.

\acknowledgments We wish to thank Gabriele Veneziano and Jacek Wosiek
for valuable discussions.

\bibliographystyle{JHEP3} 

\end{document}